\documentclass[twocolumn,secnumarabic,amssymb, superscriptaddress, nobibnotes, aps, prc]{revtex4-2}
\usepackage{mathtools}
\usepackage[percent]{overpic}
\usepackage{color}
\usepackage{soul,color}
\usepackage{contour}
\usepackage{url}
\usepackage{graphicx}
\usepackage[colorlinks=true]{hyperref}
\hypersetup{
  breaklinks=true,
  colorlinks   = true,
  urlcolor     = blue,
  linkcolor    = blue,
  citecolor   = blue
}

\begin{document}
\title{Photonuclear Cross Sections for the $^{197}$Au($\gamma$,pn)$^{195m}$Pt Reaction Near Threshold }
\author{J. Song}
\email{songj@frib.msu.edu}
\affiliation{Physics Division, Argonne National Laboratory, Lemont, IL 60439 USA}
\affiliation{Facility for Rare Isotope Beams, Michigan State University, East Lansing, MI 48824 USA}
\author{J. Nolen}
\affiliation{Physics Division, Argonne National Laboratory, Lemont, IL 60439 USA}
\author{D. Rotsch}
\affiliation{Physics Division, Argonne National Laboratory, Lemont, IL 60439 USA}
\affiliation{Radioisotope Science and Technology Division, Oak Ridge National Laboratory, Oak Ridge, TN 37830, USA}
\author{R. Gampa}
\affiliation{Physics Division, Argonne National Laboratory, Lemont, IL 60439 USA}
\author{R. M. de Kruijff}
\affiliation{Physics Division, Argonne National Laboratory, Lemont, IL 60439 USA}
\author{T. Brossard}
\affiliation{Chemical $\&$ Fuel Cycle Technologies Division, Argonne National Laboratory, Lemont, IL 60439 USA}
\author{C. R. Howell}
\affiliation{Department of Physics, Duke University, Durham, NC 27708-0308, USA}
\affiliation{Triangle Universities Nuclear Laboratory, Durham, NC 27708-0308, USA}
\author{\\F. Krishichayan}
\affiliation{Department of Physics, Duke University, Durham, NC 27708-0308, USA}
\affiliation{Triangle Universities Nuclear Laboratory, Durham, NC 27708-0308, USA}
\author{S. W. Finch}
\affiliation{Department of Physics, Duke University, Durham, NC 27708-0308, USA}
\affiliation{Triangle Universities Nuclear Laboratory, Durham, NC 27708-0308, USA}
\author{Y. K. Wu}
\affiliation{Department of Physics, Duke University, Durham, NC 27708-0308, USA}
\affiliation{Triangle Universities Nuclear Laboratory, Durham, NC 27708-0308, USA}
\author{S. Mikhailov}
\affiliation{Triangle Universities Nuclear Laboratory, Durham, NC 27708-0308, USA}
\author{M. W. Ahmed}
\affiliation{Triangle Universities Nuclear Laboratory, Durham, NC 27708-0308, USA}
\affiliation{Department of Mathematics and Physics, North Carolina Central University, Durham, NC 27707, USA}
\author{R. V. F. Janssens}
\affiliation{Triangle Universities Nuclear Laboratory, Durham, NC 27708-0308, USA}
\affiliation{Department of Physics $\&$ Astronomy, University of North Carolina at Chapel Hill, Chapel Hill, NC 27599-3255, USA}

\date{\today}

\begin{abstract}
Platinum radioisotopes are of growing interest for targeted cancer therapy and diagnostic imaging
because their decay delivers highly localized radiation doses in tissue, herewith enabling precise
DNA damage through Auger-electron emission. Developing production technologies that provide
platinum isotopes with high specific activity is 
essential in radioisotope therapy. Photonuclear reactions on
stable nuclei offer a viable accelerator-based route for isotope production when supported by reliable cross-section data.
We report photonuclear cross-section measurements for the $^{197}$Au($\gamma$,pn)$^{195m}$Pt reaction at incident $\gamma$-ray
energies of 27, 29, and 31 MeV using the activation method. The measurements were performed by
irradiating a stack of concentric-ring gold targets with a quasi-monoenergetic $\gamma$-ray beam provided by
the High Intensity Gamma-ray Source (HI$\gamma$S). The induced $^{195m}$Pt activity was quantified using
off-line $\gamma$-ray spectroscopy.
These data provide the first experimental constraints on the $^{197}$Au($\gamma$,pn)$^{195m}$Pt
cross section in the near-threshold region.
The comparison of the measured excitation function to 
PHITS and TALYS calculations indicates that the reaction becomes measurable only near 30 MeV and that
substantially higher bremsstrahlung end-point energies are required for practically meaningful production.
 \end{abstract}
\maketitle
\section{Introduction}
Platinum-195m ($^{195m}$Pt) is a metastable radionuclide with a half-life of about 4 days that decays
via isomeric transitions.
Because of its highly internally converted decay, $^{195m}$Pt emits a large number of low-energy Auger electrons,
which deposit their energy over nanometer-scale distances in the surrounding material.
Because of this feature, $^{195m}$Pt has attracted interest for targeted Auger-electron
therapy and theranostic applications~\cite{1,2}. In addition, the emission of a 98.9-keV $\gamma$-ray
makes $^{195m}$Pt suitable for quantitative SPECT imaging~\cite{3}.
The medical relevance of $^{195m}$Pt is closely linked to platinum-based chemotherapy,
where radiolabeling platinum compounds such as cisplatin enables imaging of their biodistribution
and pharmacokinetics without altering their chemical structure. Studies using $^{195m}$Pt-labeled compounds
have demonstrated quantitative SPECT imaging and highlighted its potential for imaging chemotherapy delivery
\emph{in vivo}~\citep{3,4,5}.

Production of $^{195m}$Pt has traditionally relied on reactor-based routes involving neutron capture
on enriched iridium targets~\cite{6}. Although these methods can provide substantial yields, they may be limited in achievable
specific activity due to the presence of stable platinum isotopes.
Accelerator-based production using charged particles,
such as the $^{192}$Os($\alpha$,n)$^{195m}$Pt reaction, has been shown to improve specific activity, but typically requires
enriched targets and is constrained by available beam currents~\cite{7}.
Production of $^{195m}$Pt via photonuclear reactions using electron accelerators has received comparatively
little experimental attention \cite{13}.
This approach is attractive because such reactions can produce the desired radionuclide
in a chemically distinct target material,
herewith enabling efficient chemical separation and potentially high specific activity, an important requirement
for Auger-electron-emitting
radionuclides \cite{2}.
Table \ref{table:int} summarizes the main production approaches for $^{195m}$Pt,
highlighting differences in target material, reaction channel, energy range, specific activity,
and practical limitations.

\begin{table*}
  \caption{\label{tab:dinfo}
    This table lists representative approaches for the production of $^{195m}$Pt
    highlighting their main advantages and limitations.
    The symbols in the table correspond to the following physical quantities: Resonance int. (resonance integral yield),
    $\sigma_{\mathrm{max}}$(maximum reported cross section), and EOB (end of bombardment).
  }
    \begin{ruledtabular}
    \begin{tabular}{r | l l l l l l}
      Production & Target & Reaction channel & Energy range  & Reported & Specific & Limitations  \\
      route&  & & or threshold & nuclear data  & activity &   \\
      \hline
      Reactor & $^{193}$Ir  & $^{193}$Ir(n,$\gamma$)$^{194}$Ir(n,$\gamma$)$^{195m}$Ir & reactor neutrons & Resonance int. & 38.7 GBq/g & Enriched $^{193}$Ir\\
    \cite{6}  && $\rightarrow ^{195m}$Pt & & = 2900 b & at EOB & neutron flux\\
      &&&&&& Low specific activity \\
      \hline
      Cyclotron & $^{192}$Os & $^{192}$Os($\alpha$,n)$^{195m}$Pt & 18-24 MeV & $\sigma_{\mathrm{max}}$ & $\approx$ 1 TBq/g & Enriched $^{192}$Os \\ 
      \cite{7}& & &  & = 3.7 mb & after separation & Low yield\\ 
      \hline
      Electron linac& $^{197}$Au & $^{197}$Au($\gamma$,np)$^{195m}$Pt & 13.96 MeV & $\approx$ 0.6 mb & $>$ 37 TBq/g & High energy e-linac \\
     \cite{13} & & &  & at 30 MeV & after separation & Cross section data\\
    \end{tabular}
    \end{ruledtabular}
    \label{table:int}
\end{table*}

For feasibility studies of photonuclear production of
$^{195m}$Pt, reliable cross-section data for the relevant reactions are required.
In Au-target photonuclear measurements,
the $^{197}$Au($\gamma$,n)$^{196}$Au reaction is commonly used to determine the incident photon flux
because its cross-section data are well established \citep{int_1,int_2,int_3,int_4,int_5,int_6,int_7,int_8,int_9,8}.
In constrast, the $^{197}$Au($\gamma$,np)$^{195m}$Pt reaction is the production channel of interest,
but no direct energy-dependent cross-section measurement using quasi-monoenergetic photon beams has been reported
thus far.
Although previous production of $^{195m}$Pt from gold was demonstrated using bremsstrahlung irradiation \cite{13},
such measurements provide spectrum-averaged production information rather than direct cross sections at well-defined
photon beam energies.
Cross-section measurements near the reaction threshold are, therefore, particularly important
as they define the practical onset of $^{195m}$Pt production and the minimum photon energy required
for achieving a meaningful yield.
This information is essential for optimizing photonuclear production conditions,
since irradiation below this energy would provide little production benefit.
In addition, threshold-region cross-section data provide a sensitive test of reaction-model calculations
and contribute to improving evaluated nuclear data libraries in an energy region
where the yield changes rapidly with photon energy.

In the present work, we address this data gap by measuring the $^{197}$Au($\gamma$,np)$^{195m}$Pt
cross section using quasi-monoenergetic photon beams near threshold.
The metastable isomer emits a characteristic 98.9-keV $\gamma$ ray with a half-life of 4.01 days,
enabling selective identification of the reaction channel using off-line $\gamma$-ray spectroscopy.
Photonuclear reaction rates were determined using an activation method,
as described in Sec. \ref{method:int} and \ref{expt:int}.
The measured cross sections are presented in Sec. \ref{res:int}
and the feasibility of producing $^{195m}$Pt via photonuclear reactions is discussed in Sec. \ref{con:int}.

\section{Method}
\label{method:int}

\begin{figure*}
  \begin{center}
    \begin{overpic}[width=0.95\textwidth]{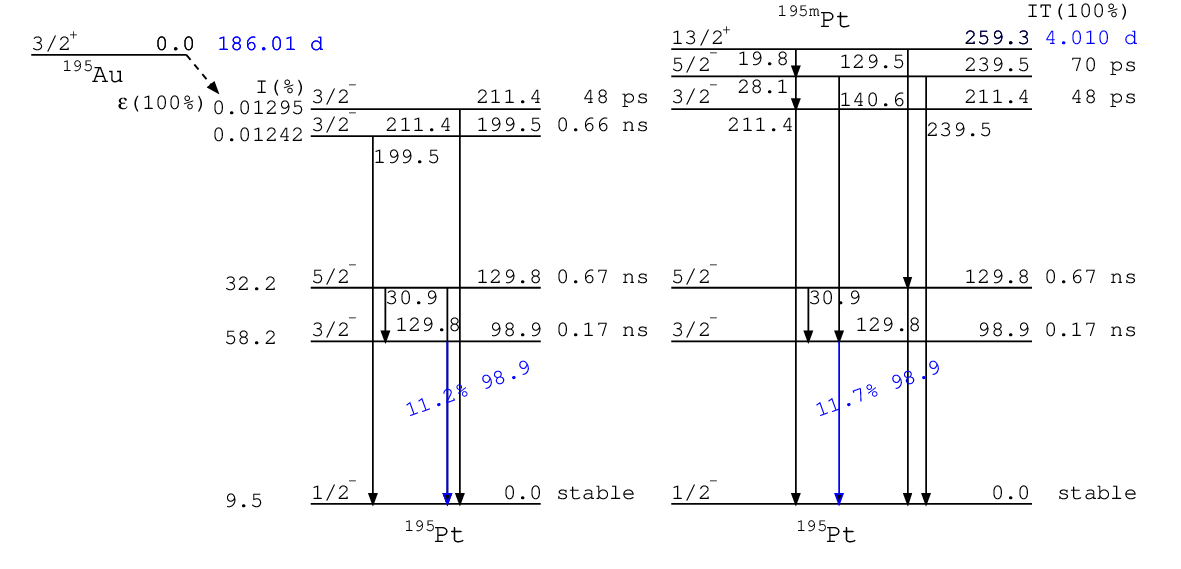}
    \end{overpic}
  \end{center}
  \caption{
    Simplified decay schemes for the $^{195}$Au and $^{195m}$Pt nuclei. The selected $\gamma$-ray
    lines (blue) were used in the analysis.
    The decay $\gamma$ energies and relative intensities are adopted from the National Nuclear Data Center (NNDC) \cite{12}.
  }
  \label{fig:0}
\end{figure*}
When a photon beam irradiates $^{197}$Au, radioactive nuclides are produced via photonuclear reactions,
including $^{197}$Au($\gamma$,pn)$^{195m}$Pt and $^{197}$Au($\gamma$,2n)$^{195}$Au,
which both decay with a branch emitting a 98.9-keV $\gamma$ ray.
Hence, the observed counts of this 98.9-keV peak contain contributions from the decays of both
$^{195m}$Pt and $^{195}$Au.
The simplified decay schemes for $^{195m}$Pt and $^{195}$Au are presented in Fig. \ref{fig:0}.
The relationship between the observed $\gamma$-ray counts and the number of
radioactive nuclei can be written as
\begin{equation}
c_{\mathrm{all}} \cdot N_0
=
n_0 e^{-\lambda t_c} \left(1 - e^{-\lambda t_m}\right),
\label{eq:1}
\end{equation}
where $N_0$ denotes the observed $\gamma$-ray counts and $n_0$ is the number of radioactive
nuclei at the end of bombardment (EOB). Here, $c_{\mathrm{all}}$ is the overall correction factor,
which  accounts for detector efficiency, detector dead time, $\gamma$ branching ratio,
$\gamma$-ray attenuation in the target, and self-absorption effects.
The symbols $\lambda$, $t_c$, and $t_m$ denote the decay constant, cooling time, and measurement time,
respectively.
Because the observed $\gamma$-ray counts represent a superposition of contributions from the two nuclides, measurements performed
at different cooling times allow the individual contributions from $^{195}$Au and $^{195m}$Pt to be disentangled.

In this work, three measurements were performed using different cooling times:
a few hours after the end of the bombardment (EOB), $\approx$ 1 week, and $\approx $ 2 weeks later.
The corrected $\gamma$-ray counts for $i$-th measurement can be expressed as
\begin{equation}
  \small
  N_{i} = \sum_{k}N_{ik} = \sum_{k} n_{0k}e^{-\lambda_{k}t_{c,i}}(1-e^{-\lambda_{k}t_{m,i}})
\label{eq:2}
\end{equation}

where $k$ labels the nuclides ($k$=1 for $^{195}$Au and $k$=2 for $^{195m}$Pt);
$N_{i}$ represents the corrected number of decay events corresponding to the observed $\gamma$-ray counts
after applying the overall correction factor $c_{all}$.
The  $t_{c,i}$ and $t_{m,i}$ symbols are the cooling and measurement times for the $i$-th measurement, respectively.
For convenience, the above expression can be rewritten in linear form as
\begin{equation}
  N_{i} = \alpha_{i}n_{01}+\beta_{i}n_{02}
\end{equation}
with \[ \alpha_{i}=e^{-\lambda_{1} t_{c,i}}(1-e^{-\lambda t_{m,i}}), \beta_{i}=e^{-\lambda_{2} t_{c,i}}(1-e^{-\lambda t_{m,i}})\]
The parameters $n_{01}$ and $n_{02}$ were determined by a weighted least-squares fit (see Appendix \ref{app:lsq} for details).
The solution can be written explicitly as
\begin{equation}
  \label{eq:7}
  \small
    \begin{pmatrix}
      n_{01} \\
      n_{02}
    \end{pmatrix}
    =
    \dfrac{1}{D}
    \begin{pmatrix}
      \sum \dfrac{\beta_{i}^{2}}{\Delta N_{i}^{2}} & -\sum \dfrac{\alpha_{i}\beta_{i}}{\Delta N_{i}^{2}}  \\
      -\sum \dfrac{\beta_{i}\alpha_{i}}{\Delta N_{i}^{2}}  & \sum \dfrac{\alpha_{i}^{2}}{\Delta N_{i}^{2}}
    \end{pmatrix}
    \begin{pmatrix}
      \sum \dfrac{\alpha_{i}}{\Delta N_{i}^{2}} N_{i} \\
      \sum \dfrac{\beta_{i}}{\Delta N_{i}^{2}} N_{i}
    \end{pmatrix}
  \end{equation}
Statistical uncertainties of $n_{01}$ and $n_{02}$ are given by
\begin{equation}
  \small
  \label{eq:10}
  \Delta n_{01} = \sqrt{\dfrac{1}{D}\sum_{i=1}^{3}\frac{\beta_{i}^{2}}{\Delta N_{i}^{2}}},
 ~ \Delta n_{02} = \sqrt{\dfrac{1}{D}\sum_{i=1}^{3}\frac{\alpha_{i}^{2}}{\Delta N_{i}^{2}}}
\end{equation}

Using Eqs. \ref{eq:7} and \ref{eq:10}, the $n_{01}$ and $n_{02}$ parameters, including associated statistical uncertainties,
can be extracted from the observed $\gamma$-ray counts, $N_{i}$.

\section{Experiment}
\label{expt:int}
A stack of five targets, each divided into three segments \cite{8},
was irradiated at seven different central photon beam energies at the High Intensity Gamma-ray Source (HI$\gamma$S) \cite{0}.
Owing to the radial
energy gradient across the target stack, the photon energy at the outermost ring was typically close to the central energy
of the subsequent lower-energy irradiation, providing an internal consistency check for the measurements
(see Ref. \cite{8} for details).
Using targets composed of three concentric segments, cross sections for the 
$^{197}$Au($\gamma$,pn)$^{195m}$Pt reaction were obtained at 3 distinct photon energies for the 31 MeV run,
as the cross sections are extremely low below 30 MeV.
Each irradiation was performed for 8–12 hours using a photon beam with fluxes in the range of 
$10^{8}$-$10^{9}$ $\gamma/s$.
The effective photon energy for each target segment was determined as the 
flux-weighted average energy of the segment-dependent photon-energy distribution 
described in Table VIII of Ref.~\cite{8}. The quoted energies correspond to these segment-averaged 
effective energies, and the horizontal error bars in Fig.~\ref{fig:2} represent the 
associated energy spread of the quasi-monoenergetic beam. The finite 
photon-energy distribution was, therefore, represented by the effective energy 
and its corresponding energy spread; no additional energy-unfolding correction 
was applied.
Detailed information on incident beam flux, irradiation time, target thickness, and beam energy
and flux for each segment is summarized in Tables III, IV, and VIII of Ref. \cite{8}.
The measurement and cooling times for 31 MeV run are summarized in Table \ref{tab:run}.

The irradiated Au target segments were measured off-line by $\gamma$-ray spectroscopy
using a calibrated high-purity germanium detector.
The samples were counted without chemical processing; each irradiated Au segment was measured in its original
metallic form after irradiation. During counting, the target segment was placed at a fixed source-to-detector
distance of 5 cm from the front face of the HPGe detector and centered on the detector axis.
The detector energy calibration was performed using standard $\gamma$-ray calibration sources
covering the energy region of interest.
The absolute full-energy peak efficiency was determined with calibrated mixed $\gamma$-ray sources
measured in the same counting geometry as the activated samples.
The resulting efficiency curve, shown in Fig.~8 of Ref.~\cite{8}, was fitted as a function
of $\gamma$-ray energy and used to determine the absolute efficiency at 98.9 keV.
Corrections were applied for detector dead time, and $\gamma$-ray attenuation/self-absorption in the Au target.
The self-attenuation correction for the 98.9-keV $\gamma$ ray was evaluated using the target thickness
and tabulated photon attenuation coefficients for Au.
The same counting geometry was used for the activated samples and calibration source,
so the geometry dependence was included in the absolute efficiency calibration.
These corrections were included in the overall correction factor used in Eq. \ref{eq:1}.

\begin{table}[h]
  \caption{\label{tab:run}Summary of counting periods. The measurement time ($t_m$) and cooling time ($t_c$)
    are given in hours for each segment.}
  \begin{ruledtabular}
    \begin{tabular}{r |r  r|r r| rr}
      &seg. 1 & & seg. 2 && seg. 3 \\
      $E_{\gamma}$ (MeV) & &  31.05 && 29.25 &&27.24 \\
      \hline
         & $t_{m}$  & $t_{c}$ & $t_{m}$ & $t_{c}$ & $t_{m}$ & $t_{c}$ \\
      run &  hr  & hr    &  hr & hr & hr & hr\\
      \hline
      1 & 5 & - & 5& - & 10 & - \\
      2 & 7 & 146.29 & 13 & 151.58 & 12 & 188.19\\
      3 & 24 & 310.59 & 18 & 332.03 & 24 & 353.69\\
    \end{tabular}
  \end{ruledtabular}
\end{table}

\begin{figure}
  \begin{center}
    \begin{overpic}[width=0.48\textwidth]{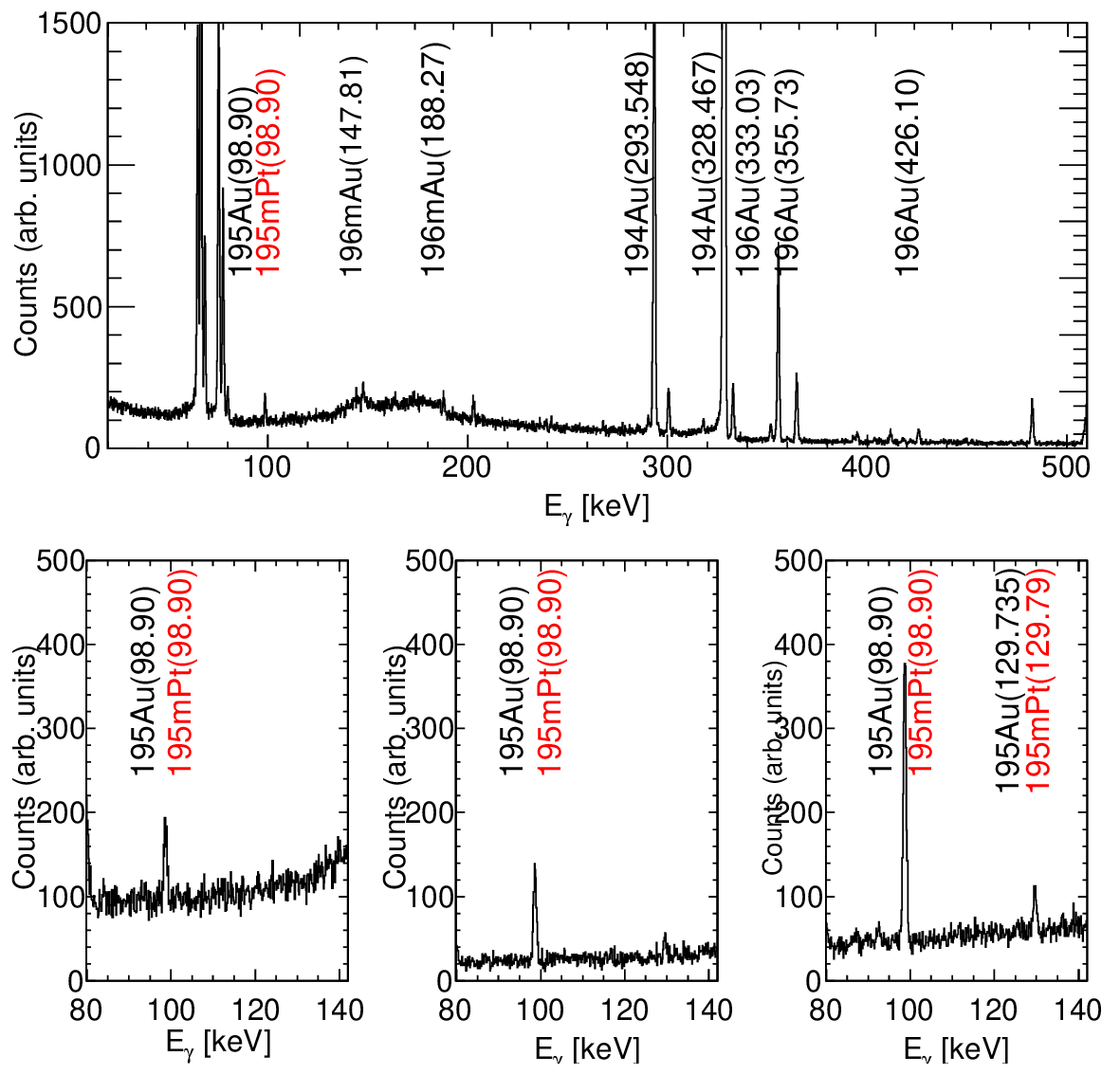}
      \put(13,89){\scriptsize a)}
      \put(8.5,42){\scriptsize b)}
      \put(41.5,42){\scriptsize c)}
      \put(74.5,42){\scriptsize d)}
      \color{red}
      \put(23.5,70){\vector(0,-1){8}}
    \end{overpic}
  \end{center}
    \caption{Full $\gamma$-ray spectrum for target segment 1 measured a few hours after the end of bombardment (EOB) (a),
      and expanded $\gamma$-ray spectra around the 98.9-keV peak measured a few hours after EOB (b), one week
      (c) and two weeks later (d).
      The 98.9-keV $\gamma$-ray line emitted by both $^{195m}$Pt and $^{195}$Au is indicated.
      The accumulated measurement times for the three measurements were 5, 7, and 24 h, respectively.}
    \label{fig:1}
\end{figure}
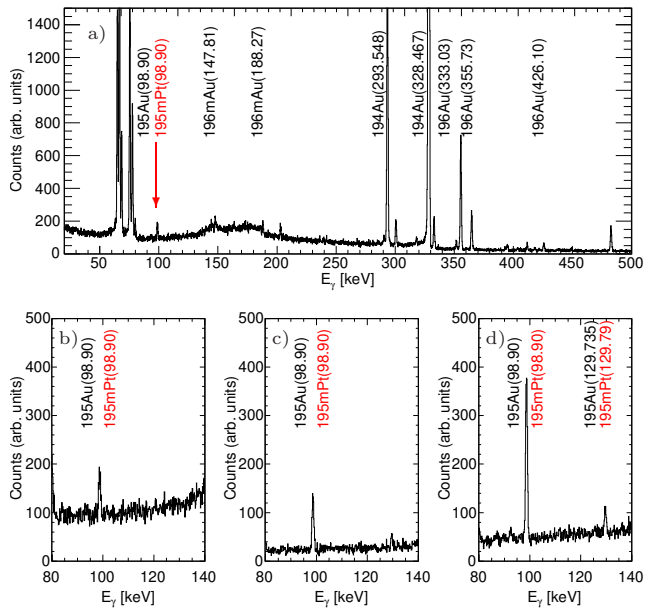

The $\gamma$-ray spectra measured for segment 1, the innermost segment,
at three different times following irradiation are presented in Fig. \ref{fig:1}:
a few hours after the end of bombardment (EOB), one week, and two weeks later.
The data accumulation times for the three measurements were 5, 7, and 24 h, respectively.
The full spectrum measured at the earliest time (see Fig. \ref{fig:1} (a)) exhibits multiple $\gamma$-ray lines originating
from activation products such as $^{196g,196m1,196m2}$Au and $^{194}$Au in the gold target.
Among these, the $\gamma$-ray line at 98.9 keV is of particular interest as it is emitted by both 
$^{195m}$Pt and $^{195}$Au whose individual contributions cannot be distinguished using only this single measurement.
The expanded spectra around the 98.9-keV region (see Figs. \ref{fig:1} (b)–(d)) demonstrate a clear evolution of the relative peak
intensities with cooling time.
At early times after irradiation, the spectrum is dominated by contributions from shorter-lived nuclides,
whereas measurements performed after longer cooling times increasingly emphasize the decay of longer-lived reaction products.
This temporal variation arises from the different half-lives of 
$^{195m}$Pt ($T_{1/2}=4.01 $ d) and $^{195}$Au ($T_{1/2} = 186.01$ d),
and enables their individual activities to be disentangled through measurements
performed at multiple cooling times.
The systematic change in the 98.9-keV peak intensity as a function of cooling time provides the experimental basis
of the approach for the
activity decomposition described in Sec. \ref{method:int}.
By combining the spectra measured at different cooling times with a least-squares analysis,
the initial activities of the contributing nuclides can be extracted reliably, despite the spectral overlap
at this 98.9-keV $\gamma$-ray energy.

\section{Results}
\label{res:int}
The photonuclear cross sections corresponding to the number of radioactive nuclei at EOB, $n_{01}$ and $n_{02}$ can be expressed as
\begin{equation}
  \label{eq:11}
  \sigma_{i} = \dfrac{\lambda_{i}n_{0i}}{N_{T}I_{0}(1-e^{-\lambda_{i}t_{0}})}
\end{equation}
Here 
$N_{T}$, $I_{0}$ and $t_{0}$ represent the number of target nuclei, incident photon flux, and irradiation time, respectively.
The statistical uncertainty is evaluated using Eq. \ref{eq:10}. Three sources of systematic uncertainty arising from
the $\gamma$-ray branching ratio, detector efficiency and beam flux are considered.
The total systematic uncertainty can therefore be expressed as
\begin{equation}
  \small
  \dfrac{\Delta \sigma}{\sigma} = \sqrt{ 
  \left(\dfrac{\Delta I_{0}}{I_{0}} \right)^{2}
    +\left(\dfrac{\Delta \epsilon}{\epsilon} \right)^{2}
    +\left(\dfrac{\Delta I_{br}}{I_{br}} \right)^{2}}
  \label{eq:12}
\end{equation}
where 
$\Delta I_{0}$, $\Delta \epsilon$, and $\Delta I_{br}$ denote the uncertainties associated with
photon beam flux, HPGe detector efficiency, and the $\gamma$ branching ratio,
respectively.
The uncertainties in the photon beam flux and HPGe detector efficiency are $\approx$ 5$\%$ and 8$\%$,
while uncertainties of the $\gamma$ branching ratios are given in Table \ref{tab:dinfo}.

\begin{table}[h]
  \caption{\label{tab:dinfo}Decay information for the measured $^{195m}$Pt and $^{195}$Au nuclei.
    Data are taken from the National Nuclear Data Center (NNDC) \citep{12}.}
  \begin{ruledtabular}
    \begin{tabular}{r | r r r r r}
      Nucl. & decay mode & half-life & $\gamma$ energy & $\gamma$ branching ratio  \\
       &  &  d & keV & \% \\
      \hline
      $^{195m}$Pt & IT 100 $\%$ & 4.010 (5) & 98.9 & 11.7 (8)\\
      $^{195}$Au & $\epsilon$ 100 $\%$ & 186.01 (6) & 98.9 & 11.21 (15)\\
    \end{tabular}
  \end{ruledtabular}
\end{table}

\begin{figure}
  \begin{center}
    \begin{overpic}[width=0.48\textwidth]{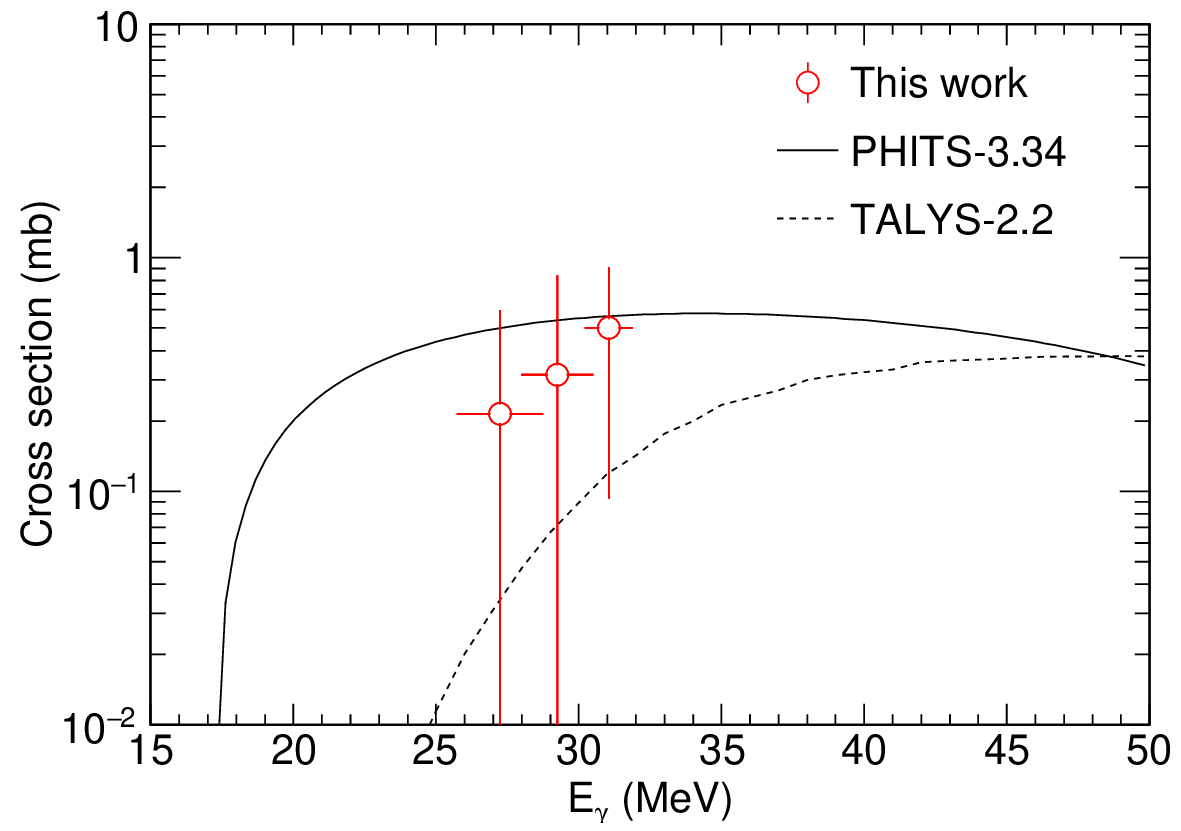}
    \end{overpic}
  \end{center}
  \caption{Excitation function of the $^{197}$Au($\gamma$,pn)$^{195m}$Pt reaction as a function of incident $\gamma$-ray energy.
    Red symbols represent the cross sections measured in this work.
    Vertical error bars indicate the total experiment uncertainties defined in Eq. \ref{eq:12},
    while horizontal error bars indicate the energy spread of the incident $\gamma$-ray beam.
    The experimental results are compared with theoretical predictions
    calculated using the PHITS-3.34 (solid black line) \cite{10} and TALYS-2.2 codes (dotted black line) \cite{11}.
  }
  \label{fig:2}
\end{figure}

The excitation function of the $^{197}$Au($\gamma$,pn)$^{195m}$Pt measured at incident $\gamma$-ray energies
of 27, 29, and 31 MeV is illustrated in Fig. \ref{fig:2}.
The measured cross sections exhibit a clear increase with photon energy,
reflecting the opening and growing contribution of the ($\gamma$,pn) reaction channel above threshold.
Vertical error bars represent the total experimental uncertainties,
which are the combined statistical (from Eq. \ref{eq:10}) and systematic (from Eq. \ref{eq:12}) uncertainties.
The statistical uncertainty of the observed $\gamma$-ray counts is the dominant contribution to the overall uncertainty
in extracting the cross sections.
This is primarily due to the relatively small contribution of the $^{195m}$Pt decay to the mixed 98.9-keV $\gamma$-ray
peak compared to that of $^{195}$Au, with the $^{195m}$Pt activity accounting for only about 5-18 $\%$ of the
total activity at EOB.
For each segment, the measured $^{195m}$Pt activity at EOB ranges from $\approx$
1.5$\times 10^{-3}$ to 4.4$\times 10^{-3}$ Bq.
As a result, the counting statistics for $^{195m}$Pt are limited and
strong anti-correlation arises between the fitted activities.
Consequently, the relative uncertainty of the extracted $^{195m}$Pt activity is comparable to,
or larger than, that of the measured cross section.

The measured cross sections are compared with theoretical predictions calculated using the PHITS \cite{10}
and TALYS \cite{11} codes.
Our data are consistent with the overall trend of both models within the experimental uncertainties.
These results provide the first experimental constraints on the cross section for the $^{197}$Au($\gamma$,pn)$^{195m}$Pt reaction in the
energy region just above reaction threshold.

\section{Conclusions}
\label{con:int}
We have measured the photonuclear cross sections for the $^{197}$Au($\gamma$,pn)$^{195m}$Pt reaction
at incident energies of 27, 29, and 31 MeV using a quasi-monoenergetic $\gamma$-ray beam at HI$\gamma$S.
The observed 98.9-keV $\gamma$-ray peak contains contributions from the decays of
both $^{195}$Au and $^{195m}$Pt. These contributions were successfully disentangled using a least-squares analysis of
measurements performed at multiple cooling times.
For the first time, cross sections were experimentally determined in the vicinity of the reaction threshold.
No statistically significant 98.9-keV peak attributable to $^{195m}$Pt was observed for irradiation energies below 27 MeV,
despite the reaction threshold energy of E$_{th}$ = 13.96 MeV, 
indicating that,
under present experimental conditions, specifically a photon flux of $\approx$ $10^{8} \gamma/s$ and irradiation time 8-10 hours,
the production yield is insufficient to detect this rare reaction channel at lower energies.
The measured cross sections indicate that the $^{197}$Au($\gamma$,pn)$^{195m}$Pt reaction begins to contribute measurably at
incident photon energies of $\approx$ 30 MeV. The data provide the first experimental constraints on the cross section in the
near-threshold region. These measured cross sections are consistent, within the experimental uncertainties, with the overall
trends predicted by the PHITS and TALYS calculations.
Although electron energies $\approx$ 30--40 MeV are commonly employed for photonuclear isotope production,
the present results suggest that such energies are insufficient to yield meaningful $^{195m}$Pt production
under realistic experimental conditions.
Combining our results with the model predictions for the excitation function for the $^{197}$Au($\gamma$,pn)$^{195m}$Pt reaction
shown in Fig. \ref{fig:2},
we project that efficient production of $^{195m}$Pt via photonuclear reactions in quantities required for applications will require
bremsstrahlung photon beams with substantially higher end-point energies than are standard, e.g., on the order of 50-60 MeV.
The present results provide new benchmark data for calculations of photonuclear production of $^{195m}$Pt
and motivate future measurements at HI$\gamma$S with higher $\gamma$-ray energies extending beyond 50 MeV.

\section*{Acknowledgments}
This research was supported by the U.S. Department of
Energy Isotope Program, managed by the Office of Science for Isotope R$\&$D and Production, under contract
DE-AC02-06CH11357 (Argonne National Laboratory),
and the Department of Energy, Office of Nuclear Physics
under grant numbers DE-SC0018112, DE-SC0018325,
DE-FG02-97ER41033 (TUNL) and DEFG02-97ER41041
(UNC). The authors would like to thank the HI$\gamma$S operators and staff for their help during the experiments.
We gratefully acknowledge the computing resources provided on Bebop, a high-performance computing cluster
operated by the Laboratory Computing Resource Center
at Argonne National Laboratory.

\appendix{}
\section{Weighted least-squares formalism}
\label{app:lsq}
From Eq.  \ref{eq:2}, the measured counts $N_{i}$ can be written as
\begin{equation*}
  N_{i} = \alpha_{i}n_{01}+\beta_{i}n_{02}
\end{equation*}
with \[ \alpha_{i}=e^{-\lambda_{1} t_{c,i}}(1-e^{-\lambda t_{m,i}}), \beta_{i}=e^{-\lambda_{2} t_{c,i}}(1-e^{-\lambda t_{m,i}})\]
The parameters $n_{01}$ and $n_{02}$ were determined by a weighted least-squares fit.
The $\chi^{2}$ is defined as
\begin{equation*}
\chi^{2}(n_{01},n_{02}) = \sum_{i=1}^{3}\dfrac{[N_{i} - (\alpha_{i}n_{01}+\beta_{i}n_{02})]^2}{\Delta N_{i}^{2}}
\end{equation*}
where $\Delta N_{i}$ denotes the statistical uncertainty of $N_{i}$.
The $\chi^{2}$ is minimized by solving,
\begin{equation*}
  \begin{aligned}
    \small
    \dfrac{\partial}{\partial n_{01}}\chi^{2} = -2\sum \dfrac{\alpha_{i}}{\Delta N_{i}^{2}} [N_{i}-(\alpha_{i}n_{01}
      +\beta_{i}n_{02})] = 0\\
      \dfrac{\partial}{\partial n_{02}}\chi^{2} = -2\sum  \dfrac{\beta_{i}}{\Delta N_{i}^{2}} [N_{i}-(\alpha_{i}n_{01}+\beta_{i}n_{02})] = 0
  \end{aligned}
\end{equation*}
Minimization of this expression leads a set of normal equations that can be written in matrix form as
\begin{equation*}
  \small
  \begin{pmatrix}
    \sum \dfrac{\alpha_{i}^{2}}{\Delta N_{i}^{2}} & \sum \dfrac{\alpha_{i}\beta_{i}}{\Delta N_{i}^{2}}
  \\
    \sum \dfrac{\beta_{i}\alpha_{i}}{\Delta N_{i}^{2}}  & \sum \dfrac{\beta_{i}^{2}}{\Delta N_{i}^{2}}
  \end{pmatrix}
  \begin{pmatrix}
    n_{01} \\
    n_{02}
  \end{pmatrix}
 =
  \begin{pmatrix}
    \sum \dfrac{\alpha_{i}}{\Delta N_{i}^{2}} N_{i} \\
    \sum \dfrac{\beta_{i}}{\Delta N_{i}^{2}} N_{i}
  \end{pmatrix}
\end{equation*}
The solution can be written explicitly as
\begin{equation}
  \label{eq:A1}
  \small
    \begin{pmatrix}
      n_{01} \\
      n_{02}
    \end{pmatrix}
    =
    \dfrac{1}{D}
    \begin{pmatrix}
      \sum \dfrac{\beta_{i}^{2}}{\Delta N_{i}^{2}} & -\sum \dfrac{\alpha_{i}\beta_{i}}{\Delta N_{i}^{2}}  \\
      -\sum \dfrac{\beta_{i}\alpha_{i}}{\Delta N_{i}^{2}}  & \sum \dfrac{\alpha_{i}^{2}}{\Delta N_{i}^{2}}
    \end{pmatrix}
    \begin{pmatrix}
      \sum \dfrac{\alpha_{i}}{\Delta N_{i}^{2}} N_{i} \\
      \sum \dfrac{\beta_{i}}{\Delta N_{i}^{2}} N_{i}
    \end{pmatrix}
  \end{equation}
where $ D = \sum\dfrac{\alpha_{i}^{2}}{\Delta N_{i}^{2}}\sum\dfrac{\beta_{i}^{2}}{\Delta N_{i}^{2}} - \left(\sum\dfrac{\alpha_{i}\beta_{i}}{\Delta N_{i}^{2}}\right)^{2}$.

The statistical uncertainty is calculated as,

\begin{equation}
  \label{eq:A2}
  \small
  \Delta n_{0k}^{2} = \sum \left[\Delta N_{i}^{2}\left( \dfrac{\partial n_{0k}}{\partial N_{i}}\right)^{2} \right]
\end{equation}
To determine the uncertainty in the paremeter $n_{01}$, we take the partial derivatives of Eq. \ref{eq:A2},
\begin{equation*}                                                                                     
  \small                                                                                              
  \dfrac{\partial n_{01}}{\partial N_{j}} = \dfrac{1}{D}\left( \dfrac{\alpha_{j}}{\Delta N_{j}^{2}}\sum\dfrac{\beta_{i}^{2}}{\Delta N_{i}^{2}}                                                              
  - \dfrac{\beta_{j}}{\Delta N_{j}^{2}}\sum\dfrac{\alpha_{i}\beta_{i}}{\Delta N_{i}^{2}}\right)       
\end{equation*}                                                                                       
and, from the Eq. \ref{eq:A1}, the statistical uncertainty of $n_{01}$ is obtained as
\begin{equation*}
  \small
  \begin{aligned}
    \Delta n_{01}^{2}
    & = \sum \left[\Delta N_{i}^{2}\left( \dfrac{\partial n_{01}}{\partial N_{i}}\right)^{2} \right]\\
    &= \dfrac{1}{D^{2}} \sum \dfrac{\beta_{i}^{2}}{\Delta N_{i}^{2}} \left[\sum\dfrac{\alpha_{j}^{2}}{\Delta N_{j}^{2}}
      \sum \dfrac{\beta_{i}^{2}}{\Delta N_{i}^{2}} - \left(\sum\dfrac{\alpha_{i}\beta_{i}}{\Delta N_{i}^{2}} \right)^{2} \right]\\
   & = \dfrac{1}{D} \sum \dfrac{\beta_{i}^{2}}{\Delta N_{i}^{2}}
  \end{aligned}
\end{equation*}
Similary, the statistical uncertainty of $n_{02}$ is obtained as,
\begin{equation*}
  \small
  \begin{aligned}
    \Delta n_{02}^{2}
    & = \sum \left[\Delta N_{i}^{2}\left( \dfrac{\partial n_{02}}{\partial N_{i}}\right)^{2} \right]\\
    &= \dfrac{1}{D^{2}} \sum \dfrac{\alpha_{i}^{2}}{\Delta N_{i}^{2}} \left[\sum\dfrac{\alpha_{j}^{2}}{\Delta N_{j}^{2}}
      \sum \dfrac{\beta_{i}^{2}}{\Delta N_{i}^{2}} - \left(\sum\dfrac{\alpha_{i}\beta_{i}}{\Delta N_{i}^{2}} \right)^{2} \right]\\
   & = \dfrac{1}{D} \sum \dfrac{\alpha_{i}^{2}}{\Delta N_{i}^{2}}
  \end{aligned}
\end{equation*}
\bibliographystyle{apsrev4-2}
\bibliography{citation}

\end{document}